\documentclass{elsart}
\usepackage[dvips]{graphicx}    
\usepackage{amssymb}

\begin{document}

\newcommand{\BQ}{\begin{equation}}
\newcommand{\EQ}{\end{equation}}
\newcommand{\BQA}{\begin{eqnarray}}
\newcommand{\EQA}{\end{eqnarray}}
\newcommand{\be}{\begin{eqnarray}}
\newcommand{\ee}{\end{eqnarray}}
\newcommand{\NN}{\nonumber \\}
\newcommand{\del}{\partial}
\newcommand{\tr}{{\rm tr}}
\newcommand{\Tr}{{\rm Tr}}
\newcommand{\Path}{{\rm P}\,}
\newcommand{\ket}[1]{\left.\left\vert #1 \right. \right\rangle}
\newcommand{\bra}[1]{\left\langle\left. #1 \right\vert\right.}
\newcommand{\ketrm}[1]{\vert {\rm #1} \rangle}  % for |phys>
\newcommand{\brarm}[1]{\langle {\rm #1} \vert}  % for <phys|
\newcommand{\V}{\widetilde V}
\newcommand{\U}{\widetilde U}
\newcommand{\x}{x_\perp}
\newcommand{\y}{y_\perp}
\newcommand{\z}{z_\perp}
\newcommand{\w}{w_\perp}
\newcommand{\q}{q_\perp}
\newcommand{\kk}{k_\perp}
\newcommand{\N}{{\mathcal N}_\tau}
\newcommand{\al}{\alpha}
\newcommand{\la}{\lambda}
\newcommand{\K}{{\mathcal K}}
\newcommand{\bb}{b_\perp}

\def\simge{\mathrel{%
   \rlap{\raise 0.511ex \hbox{$>$}}{\lower 0.511ex \hbox{$\sim$}}}}
\def\simle{\mathrel{
   \rlap{\raise 0.511ex \hbox{$<$}}{\lower 0.511ex \hbox{$\sim$}}}}

\begin{frontmatter}

\begin{flushright}
UT-Komaba/08-6
\end{flushright}

\vspace*{0.5cm}
\title{Expanding color flux tubes and instabilities}


\author{Hirotsugu Fujii}

\address{Institute of physics, University of Tokyo, Komaba, 
Tokyo, 153-8902, Japan}

\author{Kazunori Itakura}

\address{Institute of Particle and Nuclear Studies, High Energy 
Accelerator Research Organization (KEK), 1-1, Oho, Ibaraki, 
305-0801, Japan}

\begin{abstract}
We present an analytic study of the physics of the glasma 
which is a strong classical gluon field created at early stage of 
high-energy heavy-ion collisions. Our analysis is based on 
the picture that the glasma just after the collision is 
made of color electric and magnetic flux tubes extending 
in the longitudinal direction with their diameters of the 
order of $1/Q_s$ ($Q_s$ is the saturation scale of the 
colliding nuclei). We find that both the electric and 
magnetic flux tubes expand outwards and the field strength 
inside the flux tube decays rapidly in time.
Next we investigate whether there exist instabilities against small
rapidity-dependent perturbations for a fixed color configuration.
We find that the magnetic background field
exhibits an instability induced by the fluctuations in the 
lowest Landau level, and it grows exponentially in the time 
scale of $1/Q_s$. For the electric background field
we find no apparent instability while the possible relation to
the Schwinger mechanism for particle pair creations is suggested.
\end{abstract}

%\begin{keyword}
%\PACS 
%\end{keyword}
\end{frontmatter}

\section{Introduction}

There is a missing link in our current understanding of
the event evolution in ultra-relativistic heavy-ion collisions.
Before the collision, 
the projectile nuclei at very high energies are described
in the framework of the Color Glass Condensate (CGC) \cite{CGC}. 
Once the quark-gluon plasma (QGP) is formed in local thermal equilibrium,
its time evolution may be described with hydrodynamics \cite{Heinz}. 
Obviously here it is a big theoretical challenge how 
the local equilibrium state is established from the non-equilibrium 
initial condition given by the CGC. In particular, ideal
hydrodynamic simulations seem to require that, at RHIC, 
thermalization should be realized
within a very short period $\tau < 1$ fm$/c$,
which was a surprise from the viewpoint of the
standard theoretical estimates.

Theoretical efforts involve the detailed analyses of the 
plasma instability (in particular, the Weibel instability) 
\cite{Plasma_instability}
which takes place when the hard particles 
having very anisotropic distribution in the momentum space
interact with soft gauge fields (See 
Ref.~\cite{YKIS} for other scenarios and more references). 
Although many new interesting phenomena have been found in these
studies, they are based on the kinetic equation applicable for
$\tau \simge 1/Q_s$ when individual particles having transverse
momenta greater than $Q_s$ are formed out of fields. The plasma
instability scenario, therefore, should be carefully examined at
earlier stages $\tau \simle 1/Q_s$, where the system is so dense
that the field description may be more appropriate.
Yet, it is claimed that there is a certain intermediate
regime where both the Boltzmann and classical field equations
give almost the same description for the system
\cite{Mueller-Son}. 
It is therefore very likely that the plasma instability scenario
itself may be extended to such earlier times $\tau \simle 1/Q_s$
where the kinetic theory should be turned over to the classical
Yang-Mills equation; thus we shall investigate the instability
problem in the classical Yang-Mills description.

The dense pre-equilibrium system appearing between
the first impact and the equilibrated QGP
is recently named {\it Glasma} \cite{Lappi-Larry}.
The glasma, produced from the CGC initial condition,
is still well-described 
by strong coherent Yang-Mills field, which can be treated as
a weak-coupling system.  However, unique to the glasma 
is that it has strong time dependence which is believed to lead
eventually to the QGP.

If there are unstable modes in the glasma, 
it will help the matter evolve
to an isotropic/thermalized state more rapidly. 
In fact, instability in the glasma is already found in the 
lattice numerical simulation \cite{Raju}. Properties of 
the unstable glasma turned out to be similar to those of 
the Weibel instability in expanding geometry \cite{Romatschke-Rebhan}
(see also Ref.~\cite{Rebhan2} for more recent results). 
On the other hand, the instability in the glasma is 
analytically less understood  although there are 
several works about the initial fluctuation of the glasma 
\cite{Initial_fluctuation} and its possible instability 
 \cite{Muller,Kharzeev1,Kharzeev2,Fukushima,Iwazaki1}. 
In this paper, we shall present a detailed analytic study 
of the dynamics and instability of the glasma.

Our theoretical framework is based on the following picture.
First of all, we look for instabilities of the glasma 
in a single event to which one fixed color configuration is 
assigned from the initial distribution of the CGC. 
Namely we presume that thermal equilibrium should
be achieved for each event without taking the average 
over the distribution of the initial condition. The typical initial
configuration specified by the CGC is dense color charges
randomly populated on the two-dimensional transverse plane with 
its coherence length given by $1/Q_s$. After the collision,
this configuration induces strong electric and magnetic fields 
in the longitudinal direction $gE^z\sim gB^z \sim Q_s^2$, 
while the transverse components are vanishing $E^i=B^i=0$. 
Therefore, the glasma just after the collision can be characterized 
by flux tubes of strong color fields extending in the longitudinal direction
with their transverse size being $1/Q_s$. This configuration
is independent of the rapidity coordinate, provided that the glasma
is expanding longitudinally at the speed of light.
We will investigate the time evolution of this flux-tube 
configuration. Notice that this boost-invariant glasma 
cannot thermalize because it does not allow longitudinal momentum. 
It is necessary to introduce the rapidity dependence to the glasma,
which is only possible through rapidity-dependent fluctuations. 
We will analytically examine the stability of the rapidity-dependent 
fluctuations, 
and find indeed that there exists an instability in the 
magnetic background.\footnote{Similar conclusion is recently obtained
in Ref.~\cite{Iwazaki2} independently of ours.}

The present paper is organized as follows: In the next section,
we will define more precisely our theoretical framework, 
and discuss the properties of boost-invariant glasma 
as a collection of longitudinal flux tubes. Then, in Sect.~3,
we will show a detailed analysis of the rapidity-dependent 
fluctuations in the boost-invariant background.
Summary and discussion will follow in the last section.

\section{Dynamics of the boost-invariant glasma}

At high energies, heavy-ion collisions can be viewed 
as the CGC-CGC collisions \cite{Glasma_basic}. Namely, 
the incident nuclei, right-moving 1 and left-moving 2,
are highly Lorentz contracted
and the valence and large-$x$ degrees of freedom of each nucleus
can be represented as static color source, $\rho_{1,2}(x_\perp)$,
which has random distribution on the transverse plane. 
The small-$x$ gluons are treated 
as the radiation field generated by these color charges. 
Thus, the collision is described by the Yang-Mills equation 
$[D_\mu, F^{\mu\nu}]=J^\nu$ with the color current 
$J^\nu=\delta^{\nu +}\delta(x^-)\rho_1(x_\perp)
+\delta^{\nu -}\delta(x^+)\rho_2(x_\perp)$,
where $x^\pm = (x^0\pm x^3)/\sqrt2$.
Notice that the current $J^\mu$ is nonzero only on the 
light-cone axes $x^\pm=0$. Thus, the dynamics of the glasma 
in the forward light cone $x^\pm >0$ (after the collision)
is simply described by the source free Yang-Mills equation
$[D_\mu, F^{\mu\nu}]=0$ 
with appropriate boundary conditions on the 
light-cone axes. It is only through this boundary 
conditions where the information of 
the colliding CGC's enters. In fact, as we will 
discuss later, the initial condition supplied by the 
CGC is quite unique, and greatly affects 
the dynamics of the glasma. 

In order to describe the glasma expanding in the 
longitudinal direction at the speed of light, it is 
convenient to introduce the $\tau$-$\eta$ coordinates 
defined by 
$
\tau=\sqrt{2x^+ x^-}$ and 
$ 
\eta= \frac12 \ln ({x^+}/{x^-})\, . 
$
In these coordinates, the source free Yang-Mills equation 
$[D_\mu, F^{\mu\nu}]=0\, $
 is rewritten separately for $\nu=\tau,\eta,i$ as 
\BQA
&&\tau [D_\tau, \frac{1}{\tau}F_{\tau\eta}]-[D_i,F_{i\eta}]=0
\, ,\label{YM1}\\
&&{}[D_\eta, \frac{1}{\tau}F_{\tau\eta}]-[D_i, \tau F_{i\tau}]=0
\, ,\label{YM2}\\
&&\frac{1}{\tau}[D_{\tau}, \tau F_{i\tau}]
-\frac{1}{\tau^2}[D_\eta, F_{i\eta}]+[D_j, F_{ji}]=0
\, .\label{YM3}
\EQA
The second equation is the Gauss law constraint, 
which is obvious if one 
rewrites it as $[D_\eta, P^\eta]+[D_i,P^i]=0$ by using 
canonical momenta defined by 
$P^\eta \equiv 
{\delta (\tau {\mathcal L})}/{\delta (\del_\tau A_\eta )}
= \frac{1}{\tau}F_{\tau \eta}$ and 
$P^i \equiv {\delta (\tau {\mathcal L})}/{\delta (\del_\tau A_i)}
 = -\tau F_{i\tau}\, $. When we solve the Yang-Mills equation, 
we have to check if the solution indeed satisfies the Gauss law.

For later convenience, we show here the definitions of 
electric and magnetic field strength $(i, j, k =1,2,3)$:
\BQ
E^i \equiv F^{i0}\, ,\qquad B^i \equiv 
-\frac 12 \epsilon_{ijk}F^{jk} \, .
\EQ
Each component is explicitly represented with respect to 
 quantities in the $\tau$-$\eta$ coordinates as
$(i=1,2)$ 
\BQA\label{Ei}
E^i&=& - F_{i\tau} \cosh \eta 
+\frac{1}{\tau} F_{i\eta}\sinh \eta \, ,\\
E^z&=& \frac{1}{\tau}F_{\tau\eta}
\, ,\\ 
B^i&=&  \epsilon_{ij} \left(F_{j\tau}\sinh \eta 
-\frac{1}{\tau}F_{j\eta}\cosh \eta\right)\, ,\\
B^z&=& -F_{12}\, .\label{Bz}
\EQA
In the rest of this section, we discuss the dynamics of the
boost-invariant glasma paying attention to 
the characteristic properties of the CGC initial conditions. 
All the calculation is done in the Fock-Schwinger gauge 
$A_\tau=\frac{1}{\tau}(x^+A^-+x^-A^+)=0\, .$

\subsection{Boost-invariant glasma}

The glasma created in high-energy collisions expands 
in the longitudinal direction at the speed of light. 
In an idealized situation, the gauge fields after 
the collisions are independent of the rapidity coordinate 
$\eta$ \cite{Glasma_basic}:
\BQA
&&A_{\eta}={\mathcal A}_\eta=-\tau^2 \alpha(\tau,x_\perp)\, ,\\
&&A_i = {\mathcal A}_i = \alpha_i (\tau,x_\perp).
\EQA
We use calligraphic letters to represent boost invariant 
glasma which corresponds to the classical background 
fields for instability problem discussed in the next section. 
These fields satisfy the following equations
\BQA
&&\frac{1}{\tau^3} \del_\tau (\tau^3 \del_\tau \alpha ) 
- [{\mathcal D}_i, [{\mathcal D}_i, \alpha]]=0\, ,\label{first}\\
&&\frac{1}{\tau} [{\mathcal D}_i, \del_\tau \alpha_i ]
-ig\tau [\alpha, \del_\tau \alpha]=0\, ,\label{second}\\
&&\frac{1}{\tau}\del_\tau (\tau \del_\tau \alpha_i )
-ig\tau^2 [\alpha,[{\mathcal D}_i, \alpha ]]
- [{\mathcal D}_j, {\mathcal F}_{ji}]=0\, ,\label{third}
\EQA
where the covariant derivative ${\mathcal D}_i$ and the field 
strength ${\mathcal F}_{ji}$ are, respectively, given by
$ [{\mathcal D}_i, \alpha] = 
\del_i \alpha -ig [\alpha_i, \alpha]$ and 
${\mathcal F}_{ji}=\del_j \alpha_i -\del_i \alpha_j 
-ig [\alpha_j, \alpha_i]\, .$

These equations are to be solved with the following 
initial condition
\BQA
&&\alpha(\tau=0,x_\perp)
=-\frac{ig}{2}[\alpha_1^i(x_\perp), \alpha_2^i(x_\perp)]\, , 
\label{IC1}\\
&&\alpha^i (\tau=0,x_\perp)
= \alpha_1^i(x_\perp)+ \alpha_2^i(x_\perp)\, ,
\label{IC2}
\EQA
where $\alpha_{1}^i(x_\perp)$ and $\alpha_{2}^i(x_\perp)$ are the 
classical gauge fields of colliding nuclei, 1~and~2, before the collision. 
Those are associated with the random color sources on $x^\pm=0$,
and therefore the initial fields $\alpha(\tau=0,x_\perp)$ and 
$\alpha_i(\tau=0,x_\perp)$, too, may be considered as random variables.
In the original McLerran-Venugopalan model, the color charge distribution 
on the transverse plane is Gaussian random and there is no 
correlation between the charges at different transverse points.
However, in reality, the charge distribution is not completely 
random. There is some structure in it. For example, it must be 
color neutral if it is averaged over the transverse area comparable with the 
nucleon radius. In fact, one can show that color neutrality 
is ensured for the transverse area whose extent is 
larger than $1/Q_s$ \cite{color_neutral}. In other words, 
this implies that the 
classical gauge field is randomly distributed on the transverse 
plane, but its coherence length is at most of the order of $1/Q_s$.

Just after the collision, only the longitudinal components of 
the electric and magnetic fields are nonzero \cite{tau-exp,Lappi-Larry}
\BQA
E^z\vert_{\tau=0^+}  &=& -ig [\alpha_1^i, \alpha_2^i]\, ,
\label{CGC_ICe}
\\
B^z\vert_{\tau=0^+} &=& ig \epsilon_{ij}[\alpha_1^i, \alpha_2^j]\, .
\label{CGC_ICb}
\EQA
Note that these fields are {\it rapidity  independent} as the gauge 
fields $\alpha_{1,2}^i(x_\perp)$ of the nuclei are assumed 
to be rapidity independent.  
On the other hand, as mentioned above, 
the coherence length in the transverse plane 
is $\sim 1/Q_s$. Therefore, one finds that the initial
configuration of the glasma gauge fields is represented as 
collection of color 
electric and magnetic flux tubes in the longitudinal direction 
with their diameters of the order of $1/Q_s$ as shown in Fig.~1.
According to the numerical results by Lappi and McLerran 
\cite{Lappi-Larry}, 
these longitudinal fields decay rapidly
while the other components are generated as $\tau$ proceeds.

In the following, for the sake of clarity,
we separately discuss two cases with 
(i) $\alpha= 0$ and $\alpha_i\neq 0$,  and with 
(ii) $\alpha\neq 0$ and $\alpha_i=0$,
which correspond to the magnetic and electric flux tubes, respectively.
In fact, the results are valid even when both
$\alpha$ and $\alpha_i$ are nonzero, provided that 
they have the same color.

\begin{figure}[t]
\begin{center}
\includegraphics[scale=0.5]{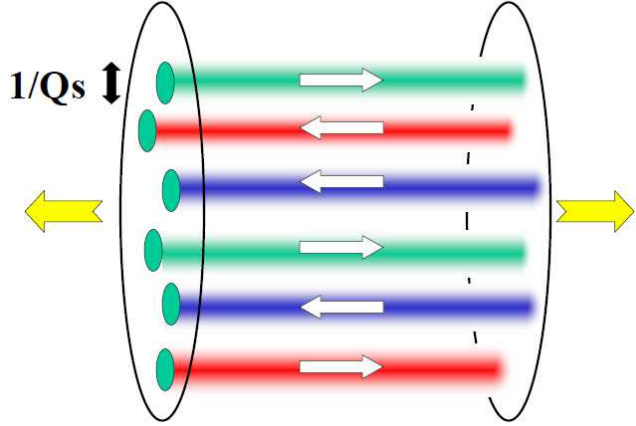}
\caption{The color electric and 
magnetic flux tubes just after the collision. 
The transverse size of the flux tube is of the order of $1/Q_s$. }
\label{flux_tube}
\end{center}
\end{figure}

\subsection{Magnetic flux tube}\label{Boost_inv_B}
Let us first consider the dynamics of a magnetic flux tube,
which is a unique object in the glasma.
We assume that the magnetic flux tube is isolated and there
is no longitudinal electric field around it.
This situation is simply realized, from 
Eqs.~(\ref{CGC_ICe}) and (\ref{CGC_ICb}), when
$\alpha_1^i$ and $\alpha_2^i$ have the same color but
$\alpha_1^i$ and $\alpha_2^j$ $(i\neq j)$ have different colors from
each other, which corresponds to the initial condition 
$\alpha(\tau=0)=0$ and $\alpha^i(\tau=0)\neq 0$ 
in Eqs.~(\ref{IC1}) and (\ref{IC2}). 
For $\tau >0$, we preserve this condition 
\BQ
\alpha = 0,\qquad \alpha_i \neq 0\, .
\EQ
Then we have the field strengths:
\BQ \label{strength_magnetic}
E^z=0\, ,\quad B^z=-F_{12}\, ,\quad
E^i= - F_{i\tau} \cosh \eta\, ,\quad
B^i= \epsilon^{ij} F_{j\tau} \sinh \eta.
\label{B_tube_general}
\EQ
Note that the transverse components $E^i$ and $B^i$ are transformed
among them under the longitudinal boost.
We consider the fields only in the local rest-frame ($\eta=0$),
and then $E^i(\eta=0)=\del_\tau \alpha_i\, ,\ B^i(\eta=0)=0.$

Notice that, by using residual gauge freedom,
one can always set $[\alpha_x, \alpha_y]=0$ so that 
the magnetic field 
$B^z=-F_{12}=-(\del_x \alpha_y - \del_y \alpha_x)$ has the same 
color structure as $\alpha_i$.
Let us further assume, as a special case, that 
$\alpha_i$ and $B^z$ are directed to a certain color.
In other words, we assume here that the initial conditions 
(\ref{IC1}) and (\ref{IC2})
allow the gauge field configurations which can be
transformed to the Abelian form with a constant color direction%
\footnote{
Incidentally, it is known that the covariantly
constant field solution to the source-free Yang-Mills equation is
generally written in a color factorized form:
$F_{\mu\nu}= \langle F_{\mu\nu} \rangle  n^a t^a$ and
$A_\mu = \langle  F_{\mu\nu} \rangle x^\nu n^a t^a$ 
where
$\langle F_{\mu\nu} \rangle$ is independent of
$x_\mu$ and $n^a$ is a constant color vector\cite{Savvidy}. 
This situation should be approximately
realized inside of the flux tube.
}.
This restricts the dynamics to the Abelian subgroup,
and one can ignore all the commutators in the equation for 
$\alpha_i$.

Since $\alpha=0$, the first Yang-Mills equation (\ref{first}) 
is trivial,
and the other two equations (\ref{second}) and (\ref{third}), 
respectively, reduce to 
\BQA
&&\del_\tau k_i \tilde \alpha_i =0\, ,\label{eq1}\\
&&\del_\tau^2 \tilde \alpha_i + \frac{1}{\tau} \del_\tau \tilde \alpha_i 
+ \kk^2 \left( \delta_{ij}-\frac{k_i k_j}{\kk^2} \right)\tilde \alpha_j =0\, ,\label{eq2}
\EQA
where we have already introduced the Fourier mode 
$\tilde \alpha_i(\tau,\kk)$ with respect to transverse coordinates. 
Note that the factor $( \delta_{ij}-{k_i k_j}/{\kk^2} )$ is a projection
operator to the direction perpendicular to $k_i$. Thus, it is convenient 
to decompose a two-dimensional vector $\tilde \alpha_i$ into 
two parts (parallel and perpendicular to $k_i$)
\BQ
\tilde \alpha_i = k_i C_{\parallel} + \epsilon_{ij}k_j C_\perp \, . 
\EQ
Substituting this into Eq.~(\ref{eq1}), one finds that $C_{\parallel}$ 
is independent of $\tau$. Indeed, this first term is the remaining gauge
freedom and unphysical. We set $C_{\parallel}=0$. 
Then the Gauss law (\ref{eq1}) 
is automatically satisfied and
Eq.~(\ref{eq2}) becomes a simpler equation for $C_\perp$:
\BQ
\del_\tau^2  C_\perp + \frac{1}{\tau} \del_\tau C_\perp + \kk^2
C_\perp=0\, ,
\label{C_perp}
\EQ 
which is solved by the Bessel function $J_0(|\kk| \tau)\, . $ 
Therefore, the solution to Eq.~(\ref{third}) with $\alpha=0$ which 
satisfies the initial condition $\alpha_i^{\rm init}(\x)$ is given by 
\BQ
\alpha_i(\tau,\x)=\int \frac{d^2 \kk }{(2\pi)^2}\, 
{\rm e}^{i\kk \x} J_0(|\kk|\tau) \ 
\tilde \alpha_i^{\rm init}(\kk)\, ,\label{solution_B}
\EQ
where the Fourier transform of the initial condition 
 must have the following form
\BQ
\tilde \alpha_i^{\rm init} (\kk) 
\equiv \int d^2x_\perp\, {\rm e}^{-ik_\perp^i x_\perp^i}
\, \alpha_i^{\rm init}(x_\perp)= \epsilon_{ij} k_j f(\kk)\, .
\EQ
Notice that the series expansion of $J_0(|k_\perp|\tau)$ 
has  only terms with even powers of $\tau$. 
This is consistent with the result of $\tau$-expansion developed in 
Ref.~\cite{tau-exp}. It is also important to notice that the 
Bessel function $J_0(z)$ is an oscillating function with its 
amplitude decreasing. 
Hence, the dynamics of the boost-invariant 
magnetic field is stable unless one allows for rapidity-dependent 
fluctuations. 
The dissipative behavior of 
the solution is understood 
as the effects of longitudinal expansion of the system.
If the second term in Eq.~(\ref{C_perp}) were absent, the solution would
be simply given by a non-dissipative oscillation ${\rm e}^{i |\kk|\tau}$. 
Thus the dissipative behavior is a result of the second term, which 
is present due to the $\tau$-$\eta$ coordinates that are suitable for 
describing expanding systems.

To get an intuitive picture of the time dependence of the solution 
(\ref{solution_B}), 
let us discuss a simplest but physically reasonable setting
for the initial condition. 
Consider a single isolated magnetic flux tube
which is represented by a Gaussian\footnote{We can perform similar 
analyses for Lorentzian profile $1/(k_\perp^2 + Q_s^2)$ in the momentum 
space.} 
with its transverse size being $1/Q_s$. 
This can be realized by the following initial condition:
\BQ
\tilde \alpha_i^{\rm init}(\kk)\propto -i 
\frac{\epsilon_{ij}k_j}{\kk^2}\, {\rm e}^{-\frac{k_\perp^2}{4Q_s^2}}\, .
\EQ
In fact, one can compute $B^z$ for $\tau > 0$ with the result
\BQ
B^z(\tau,r) 
=B_0 \, {\rm e}^{-Q_s^2r^2} {\rm e}^{-Q_s^2\tau^2 } I_0(2Q_s^2 r \tau )\, ,
\EQ 
where $r=|x_\perp|$ and $I_0(z)$ is the modified Bessel function. 
In this expression one can easily check that $B^z(\tau=0, r)$ is 
the Gaussian. 
Similarly, one can compute the transverse electric field 
$E^T\equiv \sqrt{(E^x)^2+(E^y)^2}$ as 
\BQ
E^T(\tau,r)=B_0\, {\rm e}^{-Q_s^2r^2} {\rm e}^{-Q_s^2\tau^2 }\, 
I_1(2Q_s^2 r \tau )\, .
\EQ
In Fig.~2, we show the transverse profile of $B^z$ and $E^T$ 
at five different instants $Q_s \tau=0,$ 0.5, 1.0, 1.5, 2.0. 
Remarkably, as $\tau$ increases, the magnetic flux tube expands 
outwards while the strength inside the flux tube decays rapidly. 
On the other hand, the transverse component of the electric field 
is initially zero, but grows until $Q_s \tau\sim 1$ and then 
turns to slow decrease.

All these features are intuitively understood
by looking at the evolution of the flux tube 
as a function of $t=x^0$, instead of $\tau$.
In Fig.~3 shown are the lines of
the magnetic field in the $z$-$x_\perp$ plane 
at two different instants $Q_s t =1$ (solid) 
and $2$ (dashed). 
The field strengths at $Q_s z = \pm 1$ ($\pm 2$) for $Q_s t =1$ (2)
have the Gaussian profile of the initial condition. 
Non-zero transverse magnetic field appears at $z\ne 0$,
according to Eq.~(\ref{B_tube_general}).
As the nuclei recede from each other, the flux tube in between 
is longitudinally stretched out and transversely expands,
and thence the field strength decays rapidly in the mid-rapidity region.
Notice that this
behavior is completely the same as that of ordinary electric field 
since we have assumed that the gauge field has a single color component.

\begin{figure}[t]
{\large \hspace*{0.5cm}$B^z\, (E^z)$ \hspace{4.5cm} $E^T\, (B^T)$}
\begin{center}
\includegraphics[scale=0.83]{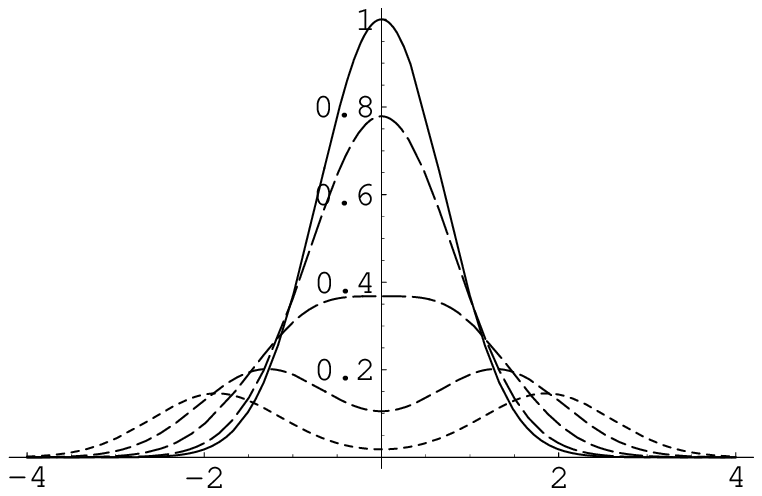}
\includegraphics[scale=0.83]{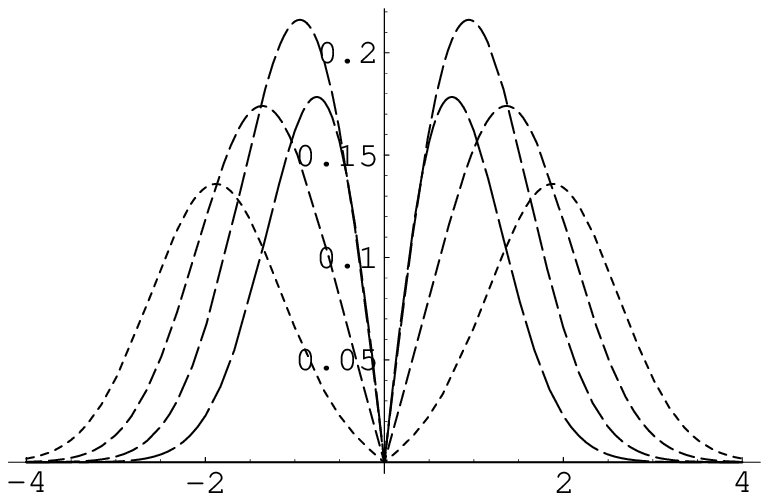}
\hspace{5cm}$Q_s x_\perp$\hspace{6.cm}$Q_s x_\perp$
\caption{Spatial profile of the longitudinal electric field 
$B^z$ (left) and the transverse magnetic field 
$E^T\equiv \sqrt{(E^x)^2+(E^y)^2}$ (right) 
at five different times $Q_s \tau  = 0 $ (solid),
\ 0.5 (longest dash),\ 1.0,\ 1.5,\ 2.0 (shortest dash). 
The maximum strength of the electric field  is normalized 
to unity at $\tau=0$, i.e., $B_0=1$. 
Notice that $E^T=0$ at $\tau=0$. These are true for 
$E^z$ (left) and $B^T$ (right) of a single electric flux tube.}
\vspace{5mm}
\label{flux_tube_EB}
\end{center}
\end{figure}

\begin{figure}[t]
\begin{center}
\includegraphics[scale=1.]{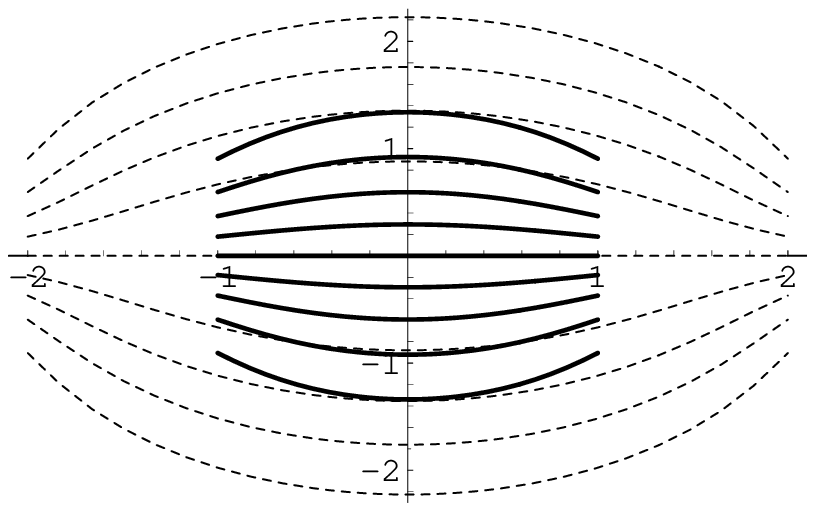}\\
 \vspace{-3cm}\hspace{9.3cm}$Q_s z$ \\
\vspace{2.5cm}
$Q_s x_\perp $
\caption{Lines of the magnetic force at $Q_s t =1$ (solid) 
and 2 (dashed). 
The sources are located at the longitudinal positions, 
$Q_s z = \pm 1$ and $\pm 2$, at $Q_s t =1$ and 2, respectively.
These can be also regarded as the lines of the electric force 
in the electric flux tube.}
\label{field_force_line}
\vspace{5mm}
\end{center}
\end{figure}

It is interesting to note that the rapid decay of 
the longitudinal magnetic field 
and the slow growth of the transverse electric field is qualitatively 
similar to the numerical result reported in Ref.~\cite{Lappi-Larry}. 
In order to confirm this similarity in a more direct way, let us take 
the average of $(B^z)^2$ and $(E^T)^2$ over the transverse plane. 
For a single isolated flux tube given by Eq.~(\ref{init_E}), 
the integration over the transverse plane gives finite results: 
\BQA\label{averageB}
\overline{ (B^z)^2 }\equiv \int d^2 x_\perp (B^z)^2
=\frac{\pi}{2}\, \frac{B_0^2}{Q_s^2}
\, {\rm e}^{-Q_s^2 \tau^2} I_0(Q_s^2 \tau^2)\, ,\\
\overline{ (E^T)^2 }\equiv \int d^2 x_\perp (E^T)^2
=\frac{\pi}{2}\, \frac{B_0^2}{Q_s^2}
\, {\rm e}^{-Q_s^2 \tau^2} I_1(Q_s^2 \tau^2)\, .
\EQA

\begin{figure}[t]
\begin{center}
\includegraphics[scale=1.]{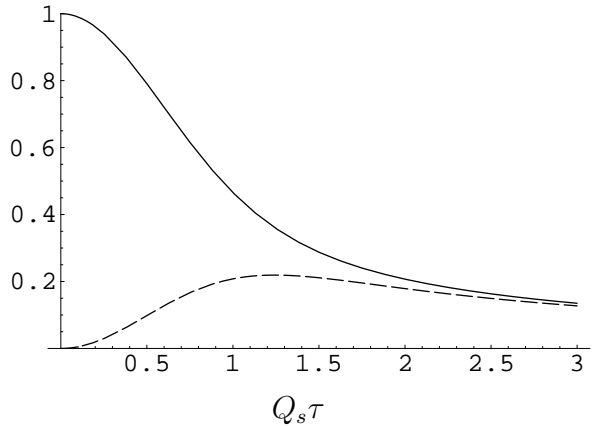} \\
$Q_s\tau$
\caption{Time dependence of the magnetic field $(B^z)^2$ (solid) and 
the transverse electric field $(E^T)^2$ (dashed) of a single 
magnetic flux tube. Both are averaged over the transverse plane, 
and normalized for $(B^z)^2$ at $\tau=0$. These are true for 
$(E^z)^2$ (solid) and $(B^T)^2$ (dashed) of a single electric
 flux tube.}
\label{time_dep_EvsB}
\end{center}
\end{figure}

In Fig.~4, we show the time dependence of $\overline{(B^z)^2}$ 
and $\overline{(E^T)^2}$ which is normalized for the magnetic 
field at $\tau=0$. This is qualitatively very similar 
to the numerical result of Ref.~\cite{Lappi-Larry}\footnote{
There is no divergence in our computation since we use
a Gaussian profile in the transverse plane for the flux tube
\cite{Fukushima}.}. 
For more realistic initial configurations with many flux tubes 
(of electric and magnetic types), 
we have to sum up all those contributions.
However, if the flux tubes are distributed homogeneously 
on the transverse plane, which will be realized for the 
initial condition of well-saturated nuclei, we can expect 
that the contribution from a single 
flux tube gives a fair approximation to the actual behavior
of $(B^z)^2$ and $(E^T)^2$ averaged over the transverse plane. 
Hence it makes sense (at least qualitatively) to compare our 
results with the numerical results of Ref.~\cite{Lappi-Larry}. 

The large-$\tau$ behavior of $(B^z)^2$ and $(E^T)^2$ can 
be easily determined from the asymptotic form of the modified Bessel 
function:
\BQ
I_\nu (z) \sim \frac{{\rm e}^z}{\sqrt{2\pi z}}\, ,\qquad  z\to \infty\, .
\label{asymptotic_Inu}
\EQ
Namely, both $(B^z)^2$ and $(E^T)^2$ show the same behavior as is 
evident from the figure:
\BQ
\overline{ (B^z)^2 }\sim \overline{ (E^T)^2 }\sim  \frac{1}{\tau}\, .
\label{large_tau}
\EQ
This is consistent with the scaling behavior of the energy density 
at large $\tau$, which is a result of the longitudinal expansion of 
the system \cite{Fukushima}. Indeed, as we will show later for the 
electric flux tubes, the behavior (\ref{large_tau}) is common 
for the longitudinal electric field and transverse magnetic field.

These time dependences of 
$\overline{ (B^z)^2 }$ and $\overline{ (E^T)^2 }$ 
are also understood from the viewpoint of the energy conservation.
The problem can be treated just as in the case of the 1-dimensional
Bjorken expansion 
since we have integrated the density over the transverse plane. 
Change of the energy density per unit rapidity is caused by 
the $pdV$ work\cite{Gyulassy-Matsui}, 
where $p$ is the longitudinal pressure and $dV$ 
is the volume change $dV=d\tau $ per unit rapidity for a 
boost-invariant expanding system. In the magnetic flux tube 
configuration, the energy density
and the longitudinal pressure \cite{Raju}
integrated over $x_\perp$ are,
respectively, given by
$\frac{\tau}{2}[\overline{ (E^T)^2 } +\overline{ (B^z)^2 }]$ 
and
$\frac{1}{2}[\overline{ (E^T)^2 } - \overline{ (B^z)^2 }]$ at $\eta=0$. 
Then the energy conservation implies
\BQ
d \left (
\frac{\tau}{2}
\left [ \, 
\overline{ (E^T)^2 } +\overline{ (B^z)^2 }
\, \right ]
\right ) 
= -
\frac{1}{2} 
\left [ \, 
\overline{ (E^T)^2 } - \overline{ (B^z)^2 }
\, \right ] \, d\tau
\, .
\EQ
As is well known, 
without the pressure term,
the energy density per unit rapidity would decrease as $1/\tau$.
However, 
for the initial configuration having the longitudinal field alone, 
the right-hand side is positive \cite{Raju,Lappi-Larry},
which slows down the decrease of the energy density.
As the transverse field $\overline{ (E^T)^2 }$ is generated, 
the longitudinal pressure is weakened
and the behavior of the energy density approaches the $1/\tau$ dependence.

\subsection{Electric flux tube}\label{Boost_inv_E}

Let us now consider the opposite situation where only a longitudinal
electric field is present. 
This situation is simply realized, from 
Eqs.~(\ref{CGC_ICe}) and (\ref{CGC_ICb}),
when
$\alpha_1^i$ and $\alpha_2^j$  $(i\neq j)$ have the same color
but $\alpha_1^i$ and $\alpha_2^i$ have different colors from each
other.
In this case, $\alpha_i$ is pure-gauge in the transverse plane 
and can be eliminated by a gauge transformation.
Therefore we consider the configuration
\BQ \label{electric}
\alpha\neq 0,\qquad \alpha_i=0. 
\EQ
Then, each component of the field strength is given by (at $\eta=0$) 
\BQA
&&E^i = 0\, ,\qquad\quad\ \, 
E^z =\frac{1}{\tau}\del_\tau (-\tau^2 \alpha)\, ,\\
&&B^i=\tau \epsilon_{ij}\del_j \alpha\, ,\quad  B^z=0\, .
\EQA
We again assume that $\alpha$ is oriented to a certain color 
direction.
This color direction of the isolated electric flux tube is
independent of that of $\alpha^i$ for the magnetic flux tube 
in the previous subsection. 
While the second 
and third Yang-Mills equations, (\ref{second}) and (\ref{third}), 
are automatically satisfied under this assumption, 
the first equation (\ref{first}) reduces to a very simple 
differential equation:
\BQ\label{diff_E}
\frac{1}{\tau^3}\del_\tau (\tau^3 \del_\tau \alpha)-\del_i^2 \alpha=0\, .
\EQ
Performing the Fourier transform in the transverse space, one 
finds that this equation is identical with the Bessel equation. 
Therefore, the solution satisfying the initial condition 
$\alpha(\tau=0,x_\perp)=\alpha_{\rm init}(x_\perp)$ is given by 
\BQ\label{solution_E}
\alpha(\tau,x_\perp)=\int \frac{d^2 k_\perp}{(2\pi)^2} \, 
{\rm e}^{ik_\perp x_\perp} 
\frac{2}{|k_\perp|\tau} J_1(|k_\perp|\tau)\, 
 \tilde \alpha_{\rm init}(k_\perp)  \, ,
\EQ 
where $\tilde \alpha_{\rm init}(\kk)$ is the Fourier transform of the 
initial condition $\alpha_{\rm init}(x_\perp)$.
One can easily check that this solution satisfies the initial condition
by using 
$\lim_{\tau\to 0}\frac{2 }{|k_\perp|\tau} J_1(|k_\perp|\tau) = 1.$ 
Similarly as in the case of the magnetic flux tubes, 
the expansion of $J_1(|k_\perp|\tau)/|k_\perp|\tau$ 
with respect to $\tau$ gives only even powers of $\tau$, which 
is consistent with the result of $\tau$-expansion developed in 
Ref.~\cite{tau-exp}. In addition to this, we note that $J_1(z)/z$ 
is an oscillating function with its amplitude decreasing.
Thus, the dynamics of the boost-invariant background electric 
field is stable, too.

Let us again consider a simple example in order to understand 
the time dependence of the solution (\ref{solution_E}).
Consider a single isolated electric flux tube 
whose transverse size is of the order of $1/Q_s$. 
This can be represented by the initial condition of the Gaussian form:
\BQ \label{init_E}
\alpha_{\rm init}(\x) = \alpha_0 \, {\rm e}^{-\x^2 Q_s^2}\, .
\EQ
Inserting (the Fourier transform of) this into Eq.~(\ref{solution_E}), 
one finds that the result is represented by infinite series of 
hypergeometric functions. However, the expressions for the longitudinal 
electric field $E^z$ and the transverse magnetic field 
$B^T\equiv \sqrt{(B^x)^2+(B^y)^2}$ become remarkably simple: 
\BQ
E^z(\tau, r)=E_0 \, 
{\rm e}^{-Q_s^2 r^2} {\rm e}^{-Q_s^2 \tau^2 }\, I_0(2Q_s^2 r \tau )\, ,
\EQ
\BQ
B^T(\tau,r)=E_0\, 
{\rm e}^{-Q_s^2 r^2} {\rm e}^{-Q_s^2 \tau^2 }\, I_1(2Q_s^2 r \tau )\, .
\EQ
Notice that these are the same functional form as 
$B^z$ and $E^T$ in the previous subsection. 
Thus we can simply replace all the results discussed for 
$B^z$ and $E^T$ by $E^z$ and $B^T$. In particular, 
Figures~2 and 4 are equally true for $E^z$ and $B^T$, 
and Figure 3 is true for the electric field.

\subsection{Coexisting case}

So far, we have discussed magnetic and electric flux tubes separately.
However, in fact, if both $\alpha$ and $\alpha_i$ are oriented to 
the same direction in the color space, then one can ignore all 
the commutators in the Yang-Mills equations (\ref{first})--(\ref{third}) 
yielding decoupled equations for $\alpha_i$  and $\alpha$.
Namely, they are given by Eq.~(\ref{eq2}) and Eq.~(\ref{diff_E}), 
respectively. 
Moreover, as is evident from the explicit form of the field strength 
(\ref{Ei})--(\ref{Bz}), the contributions to the field strength from $\alpha$ 
and $\alpha_i$ are additive. Namely, for $\eta=0$, one obtains 
$E^i=\del_\tau \alpha_i$, $E^z=\frac{1}{\tau}\del_\tau(-\tau^2 \alpha)$,
$B^i=\tau \epsilon_{ij}\del_j \alpha$, and 
$B^z=-\del_x \alpha_y +\del_y \alpha_x$. 
Therefore, all the results in the previous 
subsections are valid even in the case where both $\alpha$ and $\alpha_i$
are nonzero as far as they have the same color structure.

\section{Instabilities of the glasma}

In the previous section, we have studied the dynamics of 
boost-invariant glasma as a collection of flux tubes
which have typical diameters of $1/Q_s$ and extend 
between the two receding nuclei. 
There we have learned that the flux tubes (longitudinal 
electric and magnetic fields) expand towards outside  
in the transverse directions, and the strength inside 
the tubes rapidly decreases. On the other hand, the 
transverse components of the field strength $B^T$ and $E^T$ 
are zero at $\tau=0$ but grow slowly until around $Q_s\tau\sim 1$, 
where the magnitudes of $B^T$ and $E^T$ become comparable with 
the longitudinal components $E^z$ and $B^z$. 
This fact, however, does not imply isotropization of the system. 
The decay of the longitudinal fields is, in our 
picture, due to the transverse expansion of the flux tube, 
which also induces nonzero transverse components of the field
strength. Thus, all these can be understood as the {\it process 
towards homogeneity in the transverse plane}. 
In fact, the gauge fields $\alpha$ and $\alpha_i$
are still {\it rapidity independent}. Therefore, thermalization 
or even isotropization of the system never occurs with 
this boost-invariant glasma. This is where the rapidity-dependent 
small fluctuations come into play. 

In this section, we 
discuss the stability of boost-invariant glasma against 
rapidity-dependent perturbations. The small fluctuations 
$a_{i,\eta}$ are defined by 
\BQA
A_i = {\mathcal A}_i +a_i(\tau,\eta,x_\perp)
\, ,\quad A_\eta = {\mathcal A}_\eta + a_\eta(\tau,\eta,x_\perp)\, ,
\EQA
where the background fields 
${\mathcal A}_i$ and ${\mathcal A}_\eta$ are boost-invariant glasma
discussed in the previous section 
(We stay in the Fock-Schwinger gauge $A_\tau={\mathcal A}_\tau + a_\tau =0$).
The stability of the boost-invariant glasma
can be examined by analyzing the Yang-Mills equations 
(\ref{YM1})--(\ref{YM3}) to the linear order in the 
fluctuations $a_i$ and $a_\eta$ on the background fields. 
The non-Abelian nature uniquely fixes the coupling
between the background field and the fluctuations.

In our calculation, to make the situation simple, 
we approximate the boost-invariant background fields by 
transversely constant electric or magnetic fields which are of course 
rapidity independent.
This simplification is a reasonable 
approximation for the fields inside the flux tube. Thus, we implicitly 
assume that the transverse extent of the constant electric or magnetic 
field is of the order of $1/Q_s$ (or slightly larger than that), 
and we will later examine if the typical transverse size of the 
unstable mode is smaller than $1/Q_s$. 
In studying the equations for the fluctuations, 
we restricted ourselves to the $SU(2)$ color gauge group and greatly 
benefited from Ref.~\cite{Chang-Weiss} for technical details. 

\subsection{Fluctuation in the magnetic background field}

\subsubsection{Constant magnetic field}

Consider the simplest case where only time-independent and 
spatially constant magnetic field is present as the background field. 
This is a special case of Sect.~\ref{Boost_inv_B}
where we treated only $\alpha_i(\tau,x_\perp)$. In the following, we 
consider color $SU(2)$ group, and assume that the constant magnetic 
field is directed to the third color component. This is realized by 
\BQ
\alpha=0,\qquad 
\alpha_i^a = \delta^{a3}\frac{B}{2}\, (y \delta_{i1}-x\delta_{i2})\, ,
\EQ
where $B(>0)$ is independent of $\tau$. 
The equations for fluctuations are coupled equations 
for $a_\eta$ and $a_i$, 
and are still complicated as they are. If one assumes that 
$a_\eta \neq 0$ and $a_i=0$, then one can show that $a_\eta$ is stable. 
We expect this is true even in the full coupled equations, 
and we rather consider the opposite case: $a_\eta=0$ and 
$a_i \neq 0\, .$ Then the Yang-Mills equations are
\BQA
&&[{\mathcal D}_i, \del_\eta a_i ]=0\, ,\\
&&{}[{\mathcal D}_i, \tau \del_\tau a_i]=0\, ,\\
&&-\frac{1}{\tau}\del_\tau (\tau \del_\tau a_i) + \frac{1}{\tau^2}\del_\eta^2 a_i
+ [{\mathcal D}_j, \delta {\mathcal F}_{ji}]-ig [a_j, {\mathcal F}_{ji}]=0\, ,
\EQA
where $\delta {\mathcal F}_{ji}=[{\mathcal D}_j, a_i]-[{\mathcal D}_i,a_j]$.
The $\eta$ dependence can be easily treated by introducing the 
Fourier transform. The first equation implies:
\BQ
[{\mathcal D}_i, a_i]=0\, . 
\EQ
This is the Gauss law. If this is satisfied, the second 
equation is trivially satisfied, too. 
The Gauss law is still very helpful in simplifying the third equation. 
After a long manipulation, we can rewrite the third equation as follows:
\BQA
&&-\frac{1}{\tau}\del_\tau (\tau \del_\tau \tilde a_i^a) - 
\frac{\nu^2}{\tau^2} \tilde a_i^a 
+ \del_\perp^2\tilde a_i^a + g B \epsilon^{3ab}\del_\theta \tilde a_i^b\NN
&&
+2g B \epsilon^{3ab} \left(\tilde a_2^b \delta_{i1}-\tilde a_1^b
\delta_{i2}\right)
+ \frac{g^2B^2 r^2}{4}\epsilon^{3cb}\epsilon^{3ba}
\tilde a_i^c=0\, ,\label{magnetic_basic_eq}
\EQA
where we have introduced the Fourier transform with respect to $\eta$
\BQ
a_i (\tau, \eta, \x) 
= \int \frac{d\nu}{2\pi}\ {\rm e}^{i\nu \eta}\ \tilde a_i (\tau, \nu, \x)\, ,
\EQ
and cylindrical coordinates 
\BQ
x=r \cos \theta, \qquad y=r \sin \theta \, .
\EQ
At this stage, it is obvious that the `neutral' 
fluctuation $a^3_i$ is stable. 
The solution of Eq.~(\ref{magnetic_basic_eq})
without the coupling to the constant $B$, 
is given by the Bessel function 
$\tilde \alpha_i^{3}\propto J_{i\nu} (|k_\perp|\tau)$, 
which shows a dissipative oscillation.

On the other hand, the other color components are not diagonal, 
and couples with each other through the magnetic field $B$. 
Physically, these fields behave as `charged' matter coupled to the 
gauge field, and show cyclotron motion. Thus it is convenient to 
introduce eigen-states of angular momentum operator 
in the transverse plane, which is given by 
the $\theta$-derivative:
\BQ
\frac{\del}{\del \theta}=x\del_2-y\del_1 \equiv -i \hat L \, .
\EQ
Taking the eigen-states of $\hat L$, i.e., 
$\tilde a^a_i\propto {\rm e}^{im\theta}$, we can replace 
$\del_\theta$ by $im$. Then we find that the transverse 
structure is equivalent to the two-dimensional harmonic oscillator.
This is clearly seen if we introduce the combined fields for colors 
\BQ
\tilde a_i^{(\pm)}=\tilde a_i^1 \pm i\, \tilde a_i^2\, , 
\EQ
and moreover for the transverse coordinate
\BQ
\tilde a_\pm^{(\cdot)}=\tilde a_1^{(\cdot)} \pm i\, \tilde a_2^{(\cdot)}\, .
\EQ
Now the equation is diagonal with respect to these fields:
\BQA
\frac{1}{\tau}\del_\tau (\tau \del_\tau \tilde a_+^{(\pm)}) 
+ \left( \frac{\nu^2}{\tau^2} \mp mg B \pm 2gB\right) \tilde a_+^{(\pm)}
+ \left(-\del_\perp^2 +\frac{g^2B^2 r^2}{4}\right) \tilde a_+^{(\pm)}
=0\, ,\NN
\frac{1}{\tau}\del_\tau (\tau \del_\tau \tilde a_-^{(\pm)}) 
+ \left( \frac{\nu^2}{\tau^2} \mp mg B \mp 2gB\right) \tilde a_-^{(\pm)}
+ \left(-\del_\perp^2 +\frac{g^2B^2 r^2}{4}\right) \tilde a_-^{(\pm)}
=0\, .\nonumber
\EQA
The last terms can be identified with the Hamiltonian of the 
two-dimensional harmonic oscillator. Therefore, one can 
replace them by the energy eigen-values $E_n a_\pm^{(\pm)}$ 
with $E_n=(2n+|m|+1)gB$. 
Then we finally obtain the equations of the familiar form:
\BQA
\frac{1}{\tau}\del_\tau (\tau \del_\tau \tilde a_+^{(\pm)}) 
+ \left(E_n  \mp mg B \pm 2gB + \frac{\nu^2}{\tau^2}\right) 
\tilde a_+^{(\pm)}=0\, ,\label{plus}\\
\frac{1}{\tau}\del_\tau (\tau \del_\tau \tilde a_-^{(\pm)}) 
+ \left(E_n  \mp mg B \mp 2gB + \frac{\nu^2}{\tau^2}\right) 
\tilde a_-^{(\pm)}=0\, .\label{minus}
\EQA
These are equations for the Bessel functions. Notice that 
the sign of $E_n  \mp mg B \pm 2gB$ or $E_n  \mp mg B \mp 2gB $ 
crucially affects the behavior of the solution. If it is 
negative, the solution is unstable. For the $+$ component 
of the spatial coordinates (Eq.~(\ref{plus})), 
it can be negative for $n=0$, color $-$, and $m\le 0$, i.e., 
$E_0  + m gB - 2gB=-gB$. 
On the other hand, for the $-$ component of the spatial 
coordinates (Eq.~(\ref{minus})), it can be negative for 
$n=0$, color $+$, and $m\ge 0$, i.e., $E_0-mgB-2gB=-gB$. 
Both cases reduce to the same equation which has an unstable 
solution:
\BQ
\del_\tau^2 f +\frac{1}{\tau}\del_\tau f
+ \left(-gB +\frac{\nu^2}{\tau^2}\right) f=0\, .
\EQ
The solution is given by the modified Bessel function:
\BQ
\tilde a^{(-)}_+,\ \tilde a^{(+)}_- \propto 
{\rm e}^{im\theta}\, r^{|m|}\, {\rm e}^{-\frac{gB r^2}{4}}\, 
I_{i\nu}\left( \sqrt{gB}\, \tau\right)\, .\label{magnetic_instability}
\EQ
These are the exact solutions to the linearized equations for 
the fluctuation $a_i(\tau,\eta,x_\perp)$ in a constant magnetic field.
The functional dependence on $m$ and $r=|x_\perp|$ comes from the 
wavefunction of the two-dimensional harmonic oscillator with 
$n=0$. The $\tau$ dependence is completely determined by the 
modified Bessel function, which is consistent with the results 
suggested in Ref.~\cite{Fukushima}.
By using the asymptotic form (\ref{asymptotic_Inu}) 
of $I_{\nu}(z)$ whose leading term is independent of $\nu$, 
we indeed find that the above solution is exponentially divergent:
\BQ
\tilde a^{(-)}_+,\ \tilde a^{(+)}_- \sim 
\frac{{\rm e}^{\sqrt{gB}\tau}}{\sqrt{2\pi \sqrt{gB}\, \tau}}\, .
\EQ
Therefore, instability exists for any $\nu$ 
although appearance of the instability depends on $\nu$, 
as we will discuss later.
All the other modes except for those mentioned above 
are stable, and their time dependence is governed by the Bessel 
function $J_{i\nu}(\sqrt{gB}\, \tau)$.

As we already mentioned, the fluctuations $\tilde a^{(\pm)}$ 
as the charged matter obey cyclotron 
motion in a uniform magnetic field.  The unstable modes appear
in the lowest Landau level ($n=0$). Interestingly,
the fluctuations $\tilde a^{(\pm)}$, having the opposite sign of
charge, show instabilities in the opposite
cyclotron rotations to each other.
In fact, the existence of 
instabilities in a constant non-Abelian magnetic field has been known
in the ordinary cartesian coordinates \cite{Chang-Weiss,Nielsen-Olesen}.
What we have done here is essentially to find the same instability 
phenomena in the $\tau$-$\eta$ coordinates 
(see Ref.~\cite{Iwazaki2} for a similar conclusion). 

Let us investigate more about the properties of unstable modes. 
First of all, the smallest spatial size of the instability is 
given by $m=0$. Namely, the Gaussian profile in the two dimensional
plane has the spatial size $L$ given by  
\BQ
L\sim \frac{1}{\sqrt{gB}}\, .
\EQ 
On the other hand, the unstable mode with $|m|\neq 0$ is 
distributed in a ring of radius $\ell \sim \sqrt{2|m|/gB}$. 
For the magnetic field created in the CGC, $gB\sim Q_s^2$, 
one finds $L\sim 1/Q_s$. Recall that the transverse size of 
the flux tube is $\sim 1/Q_s$ at $\tau=0$, but it will expand 
outward in the transverse plane. Therefore, 
we conclude that {\it the magnetic instability 
(\ref{magnetic_instability})
 will indeed be realized in the time evolution of the glasma}.

\begin{figure}[t]
\hspace*{2.5cm}$\Re {\rm e} \{I_{i\nu}(z)\}$\\
\begin{center}
\includegraphics[scale=1.]{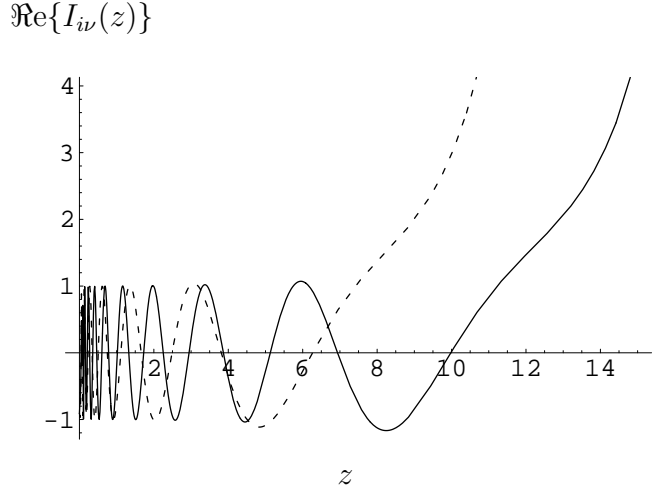} \\
$z$
\caption{$\Re {\rm e} \{I_{i\nu}(z)\}$ for $\nu=8$ (dashed) and $\nu=12$ 
(solid) normalized at $z=0.1$}
\label{modified_Bessel}
\end{center}
\end{figure} 

Next, we carefully investigate the time evolution of the 
unstable modes (\ref{magnetic_instability}) which is completely
determined by the modified Bessel function $I_{i\nu}(z)$. 
In Fig.~5, we show typical behaviors of 
$\Re {\rm e}\{ I_{i\nu}(z)\}$ for two different values of $\nu$. 
The figure indeed shows divergent behavior suggested by the 
asymptotic form (Eq.(\ref{asymptotic_Inu})). The time scale for the 
instability to grow is given by 
\BQ
\tau_{\rm grow}\sim \frac{1}{\sqrt{gB}}\sim \frac{1}{Q_s}\, .
\EQ
However, the time when the instability sets in
depends on the value of $\nu$, which we denote 
as $z_{\rm wait}$ (because we have to wait until $z=z_{\rm wait}$). 
 From Fig.~5, one may infer a rough estimate
\BQ
z_{\rm wait}\sim \nu\, .\label{z_inst}
\EQ 
This can be indeed justified in two different ways as we discuss below. 
Consider first the explicit form of $I_{i\nu}(z)$:
\BQ
I_{i\nu}(z)=\left(\frac{z}{2}\right)^{i\nu}\sum_{n=0}^\infty 
\frac{\left({z}/{2}\right)^{2n}}{n! \Gamma(i\nu + n+1)}\, .
\EQ
Notice that the term $(z/2)^{i\nu}
={\rm e}^{ i\nu \ln z/2 }$ simply oscillates for real $\nu$ and $z$,
and thus it is not relevant for the instability. One can estimate 
$z_{\rm wait}$ from the other part. Taking the first two terms 
in the infinite series, one finds
$$
I_{i\nu}(z)\sim {\rm e}^{i\nu\ln \frac{z}{2}}
\left\{\frac{1}{\Gamma(i\nu+1)}+\frac{(z/2)^2}{\Gamma(i\nu+2)}\right\}
=\frac{{\rm e}^{i\nu\ln \frac{z}{2}}}{\Gamma(i\nu+1)}
\left\{1+\frac{1-i\nu}{1+\nu^2}\left(\frac{z}{2}\right)^2\right\}\, .
$$
Therefore, one can naively define the position $z_{\rm wait}$ 
where the second term becomes the same order as 
the first one. For the real part of $I_{i\nu}(z)$, this yields 
\BQ
\frac{(z_{\rm wait}/2)^2}{1+\nu^2}=\kappa^2
\EQ
for $\kappa^2 =O(1).$ Therefore, for large $\nu$, 
$z_{\rm wait}$ grows linearly as a function of $\nu$:
\BQ
z_{\rm wait} = 2\kappa \sqrt{1+\nu^2}\sim 2\kappa \nu\, .
\EQ
Choosing $2\kappa\sim 1$ leads to the rough estimate 
(\ref{z_inst}) inferred from Fig.~5. 

Another way to estimate $z_{\rm wait}$ is based on the differential 
equation. The modified Bessel function $I_{i\nu}(\beta z)$ is a 
solution to 
$f'' +f'/z + (-\beta^2 +\nu^2/z^2)f=0$. The behavior of the 
solution changes\footnote{We thank A.~Iwazaki for this point. 
Similar argument can be found in  Ref.~\cite{Iwazaki2}.} 
depending on the sign of the coefficient of 
the last term $-\beta^2 +\nu^2/z^2$. If it is positive, 
which is realized at small $z$, the solution is oscillating. 
On the other hand, if negative, then unstable.  
The sign change takes place at $-\beta^2 +\nu^2/z^2=0$, 
yielding the same estimate for 
$z_{\rm wait}$, i.e., $z_{\rm wait}=\nu$ for $\beta=1$.

If we apply this notion to our result, we can estimate the 
time $\tau_{\rm wait}$ when the instability sets in
(which may be called ``waiting time''). 
For $z=\sqrt{gB}\, \tau \sim Q_s \tau$, 
\BQ
Q_s \tau_{\rm wait} \sim  \nu\, .
\EQ
Therefore, for a fixed value of $\nu$, the instability will exist 
only for $\tau \simge \tau_{\rm wait}$. 
In fact, this waiting time $Q_s \tau_{\rm wait}$ can be much longer 
than the growth time $Q_s \tau_{\rm grow}\sim 1$ for large $\nu$. 
Although the instability grows with the characteristic 
time given by $\tau_{\rm grow}$ once it starts, 
we have to wait until $\tau_{\rm wait}$. Therefore, the 
total time for the instability to become manifest is estimated as 
\BQ
Q_s (\tau_{\rm wait}+\tau_{\rm grow}) \sim \nu + 1\, .
\EQ
It is essentially determined by the waiting time 
$\tau_{\rm wait}$ for large $\nu$. But this time scale can be 
short for small values of $\nu$. Since the plasma 
instability scenario is based on the kinetic approach which is 
valid only for $Q_s \tau\simge 1$, we expect that the 
{\it glasma instability} will be more 
operative for the rapid thermalization of 
the matter created in the heavy-ion collisions.

The fact that the mode of $\nu$ has the ``waiting time'' 
$\tau_{\rm wait}$ for the instability implies that 
at fixed $\tau$ there exists a maximum value of $\nu$ 
which can participate in the instability, and this maximum value 
increases with increasing $\tau$. In the present case, it is 
given by 
\BQ\label{nu_max}
\nu_{\rm max}\sim {Q_s \tau}\, .
\EQ
The existence of $\nu_{\rm max}$ and its linear dependence on 
$\tau$ are consistent with the observation numerically found 
in Ref.~\cite{Raju}. But the relation between 
$\nu_{\rm max}$ and $\tau$, such as the power of $\tau$ or 
the coefficient,  crucially depends on the behavior of 
the background field as we will discuss below.

\subsubsection{Time-dependent magnetic field}

So far, the longitudinal magnetic field was assumed to be 
time-independent, but the basic equation (\ref{magnetic_basic_eq}) 
is in fact true even for time-dependent background field. 
However, we have to be careful not to replace simply the 
harmonic oscillator part by its eigenvalues when the magnetic 
field is time-dependent. This is physically because the cyclotron 
radius changes in accord with $B(\tau)$, which 
invalidates the factorization of the transverse dynamics from 
the time dependence. Still, we may assume a situation
that the time-dependence
of the magnetic field is slow compared with the characteristic 
time scale of the fluctuation (namely, the growth time of instability),
which corresponds to neglecting time derivative of $B(\tau)$. 
Thus we can recover the same equation as before except that the 
magnetic field $B$ is replaced by the time-dependent one $B(\tau)$. 
For example, the lowest Landau level yields the following equation 
\BQ\label{timedepB}
\del_\tau^2 f +\frac{1}{\tau}\del_\tau f
+ \left(-gB(\tau) +\frac{\nu^2}{\tau^2}\right) f=0\, .
\EQ
This equation can be solved for the following simple case:
\BQ
B(\tau)=B_0 \left(\frac{\tau_0}{\tau}\right)^\gamma \, .
\EQ
The solution is again given by the modified Bessel function 
(for $\gamma=2$, the solution does not show rapid growth.
 It is given by a power of $\tau$, or just oscillating):
\BQ
f(\tau)\sim 
I_{i\beta\nu} \left(\beta\sqrt{gB(\tau)}\, \tau\right)\, , 
\quad \beta=\frac{2}{2-\gamma}\, , \quad\gamma\neq 2\, .
\EQ
As we mentioned before, the relation between $\nu_{\rm max}$ and 
$\tau$ depends on the precise form of the background magnetic field:
\BQ\label{nu_max_timedep}
\nu_{\rm max} \sim \sqrt{gB(\tau)}\, \tau\, .
\EQ
Linear dependence appears only for a time-independent 
magnetic field as seen in Eq.~(\ref{nu_max}). 

For the special case of $\gamma=1$, one finds 
\BQ
f\sim I_{2i\nu}\left(2\sqrt{ gB_0\tau_0\tau}\right)
\EQ
which is similar to what is expected for an expanding system. 
On the other hand, if one takes $\gamma=1/2$ which is consistent with 
the asymptotic behavior (\ref{large_tau}), the solution is given by 
\BQ\label{mag_asympt}
f\sim I_{4i\nu/3}\left(\frac{4}{3}\sqrt{gB_0\tau_0^{1/2}}\, \tau^{3/4}\right),
\EQ
which is consistent with the result of Ref.~\cite{Iwazaki2}.

For a generic time-dependent magnetic field $B(\tau)$, 
the equation is no longer solved by simple $I_{i\nu}$'s
but must be solved numerically. Still one can infer the 
qualitative behavior of the solution from the third term of 
Eq.~(\ref{timedepB}).
Assume that $B(\tau)>0$ is slowly
decreasing from a certain finite initial value $B(\tau=0)$.
Then, the coefficient of the third term $(-gB(\tau)+\nu^2/\tau^2)$ 
is positive for small $\tau$, but will turn to be negative
at a later time. Therefore, the solution
is oscillating until $\tau=\tau_{\rm wait}$ determined by 
 $-gB(\tau_{\rm wait})+\nu^2/\tau_{\rm wait}^2=0$, and then 
starts to grow, which is qualitatively in agreement with the 
behavior of the modified Bessel function shown in Fig.~5. 

Notice that the result (\ref{nu_max_timedep}) for $\nu_{\rm max}$ 
is still valid.
In particular, if one substitutes Eq.~(\ref{averageB}) as the 
time-dependent magnetic field, one obtains 
$(\bar B\equiv B_0/2)$ 
\BQ
\nu_{\rm max}\sim \sqrt{g\bar B}\, 
\left({\rm e}^{-Q_s^2 \tau^2}I_0(Q_s^2\tau^2)\right)^{1/4}\tau \, .
\EQ
As the factor 
$\left({\rm e}^{-Q_s^2 \tau^2}I_0(Q_s^2\tau^2)\right)^{1/4}$
decreases quite slowly, 
one may observe an approximate linear increase of $\nu_{\rm max}$
in $\tau$, in accordance with the numerical result \cite{Raju}. 
But more precisely, the factor depends on $\tau$ as
$\sim 1/\tau^{1/4}$, and thus $\nu_{\rm max}\sim \tau^{3/4}$
as already suggested in Eq.~(\ref{mag_asympt}).

\subsection{Fluctuation in the electric background field}

\subsubsection{Constant electric field}

Consider the case where only time-independent and spatially 
constant longitudinal electric field is present as the background field. 
This is a special case 
of Sect.~\ref{Boost_inv_E} where we treated only $\alpha(\tau,\x)$. 
Indeed, a constant electric field $E^z=E$ is realized by the following 
gauge field:
\BQ
\alpha = -\frac12 E \, , \qquad \alpha_i=0\, .
\EQ 
Although the equations for the fluctuations $a_\eta$ and $a_i$ in 
this background are coupled equations, we can simplify them similarly 
as in the magnetic case. If one assumes that $a_\eta \neq 0$ and 
$a_i=0$, then one can show that $a_\eta$ is stable. 
We expect this is true even in the full coupled equations, 
and we consider the opposite case: $a_\eta=0$ and $a_i \neq 0\, .$
The Yang-Mills equations are now simplified to the followings:
\BQA
&&\del_i [{\mathcal D}_\eta, a_i]=0\, ,\\
&&\del_i (\tau \del_\tau a_i)=0\, ,\\
&&-\frac{1}{\tau}\del_\tau (\tau \del_\tau a_i) 
 + \frac{1}{\tau^2} [{\mathcal D}_\eta, [{\mathcal D}_\eta, a_i]]
 + \del_j (\del_j a_i - \del_i a_j)=0\, .
\EQA
In order to solve these equations, 
it is convenient to introduce the Fourier transform of $a_i$ with 
respect to $\eta$ and $\x$ 
\BQ
a_i (\tau,\eta,\x)=\int \frac{d\nu d^2\kk}{(2\pi)^3}\, 
{\rm e}^{i\nu \eta}\, {\rm e}^{i\kk \x}\, \tilde a_i(\tau,\nu,\kk),
\EQ
and its perpendicular component $\tilde b$ via
\BQ
\tilde a_i(\tau,\nu,\kk) = \epsilon_{ij}k_j\, \tilde b(\tau,\nu,\kk)\, . 
\EQ
Then, the third equation can be rewritten as 
\BQ
\frac{1}{\tau} \del_\tau (\tau \del_\tau \tilde b) 
+ \kk^2 \tilde b + \frac{1}{\tau^2} 
\left(\nu^2 \tilde b -\nu g \tau^2 [E,\tilde b] 
+\frac{g^2 \tau^4}{4}[E, [E, \tilde b]] \right)
 =0\, ,
\EQ
while the first and second Yang-Mills equations are automatically
satisfied by the introduction of $\tilde b$ (because $k_i \tilde a_i=0$).

In order to further simplify the equation, let us consider the 
$SU(2)$ case and assume that the electric field is oriented 
to the third color direction. i.e.,  $E=E^a T^a$ with $E^1=E^2=0$ and 
$E^3\equiv E$. Then, one obtains
$$
\frac{1}{\tau} \del_\tau (\tau \del_\tau \tilde b^a) 
+ \frac{1}{\tau^2} \left(\nu^2 \tilde b^a -i\nu g E \tau^2 \epsilon^{3ba} 
\tilde b^b +\frac{g^2E^2\tau^4}{4}\epsilon^{3ca}\epsilon^{3cb}\tilde b^b\right)
+ \kk^2 \tilde b^a =0\, .
$$
For the `neutral' fluctuation $a=3$, the equation does not 
depend on $E$, and the solution is given by the Bessel function 
$\tilde b^3 \propto J_{i\nu}(|\kk|\tau)$.
Therefore, there is no instability in the fluctuation to the third color 
direction. On the other hand, the equation is not diagonal for 
$a=1,2$. One can easily diagonalize it by introducing 
\BQ
\tilde b^{(\pm)}=\tilde b^{1} \pm i \, \tilde b^2\, .
\EQ
Then one finds 
\BQ\label{diff_fluc_E1}
\left(\del_\tau^2 + \frac{1}{\tau}\del_\tau\right)\tilde b^{(\pm)}
+ \left\{
\kk^2 +\frac{1}{\tau^2}\left(\nu^2 \pm {\nu gE}\tau^2 
+ \frac{g^2E^2}{4}\tau^4 \right)
\right\}\tilde b^{(\pm)} =0\, .
\EQ
Let us first ignore the third term in the bracket since it
is proportional to $\tau^4$ and is small enough compared to
the other terms at the earliest times. This approximation should 
be valid for $\tau$ satisfying 
\BQ
\tau \ll \sqrt{\frac{2\nu}{gE}}\, .
\EQ
For the electric field in the CGC $gE\sim Q_s^2$, this condition reads 
\BQ
Q_s\tau \ll \sqrt{2\nu}\, .\label{validity_E}
\EQ
Within this approximation, one obtains a simple differential 
equation:
\BQ\label{diff_fluc_E2}
\left(\del_\tau^2 + \frac{1}{\tau}\del_\tau\right)\tilde b^{(\pm)}
+ \left(
\kk^2 \pm {\nu gE} +\frac{\nu^2}{\tau^2}
\right)\tilde b^{(\pm)} =0\, .
\EQ
This is nothing but the equation for the Bessel functions.
As we discussed in the magnetic case, the behavior of the 
solution depends on the sign of $\kk^2 \pm \nu g E$. 
If we assume that both $E$ and $\nu$ are positive,  then, 
for the plus sign, it is always positive, $\kk^2+\nu gE>0,$ 
and the solution is given by $\tilde b^{(+)}
\propto J_{i\nu}(\sqrt{\kk^2+\nu g E}\, \tau)$. Therefore,
the fluctuation $\tilde b^{(+)}$ is stable. 
On the other hand, for the minus sign, if $\kk^2 - \nu g E <0$, 
then the solution is given by the modified Bessel function
\BQ
\tilde b^{(-)}(\tau,\nu,\kk) \propto
 I_{i\nu}(\sqrt{\nu g E-\kk^2}\, \tau)\, .
\EQ
As has already been seen, the modified Bessel function $I_{i\nu}(z)$ 
shows unstable behavior for {\it large value} of $z$, which corresponds
to {\it late time}. 
We shall check below whether such an instability will 
indeed manifest itself within the time specified by 
Eq.~(\ref{validity_E}). 

The condition $\kk^2 - \nu g E <0$ can be regarded as the 
condition for the existence of unstable fluctuations. 
In other words, unstable fluctuation
has transverse extent $L\sim 1/\kk$ larger than $1/\sqrt{\nu g E}$.
Recall that the transverse size of the coherent gauge field is of the 
order of $1/Q_s$.
Therefore, unstable fluctuations may exist if the transverse extent 
$L$ is smaller than $1/Q_s$:
\BQ
\frac{1}{Q_s}>\frac{1}{\sqrt{\nu g E}}\sim \frac{1}{\sqrt{\nu} Q_s}\, .
\EQ
This is satisfied if the momentum $\nu$ conjugate to rapidity is large:
\BQ
\nu \simge 1.
\EQ
Since the solution is again given by the modified Bessel function 
$I_{i\nu}(z)$, all the properties discussed for the magnetic 
instability can be applicable to the present case. 
Notice that both the waiting time 
$\tau_{\rm wait}$ 
and the growth time $\tau_{\rm grow}$ 
for the instability
depend on $\kk^2$, and that 
the shortest times are realized for $\kk=0$. Namely, for fixed $\nu$, 
the shortest waiting time and the shortest growth time are, respectively, 
given by 
\BQA
&&Q_s \tau_{\rm wait} \sim \sqrt{\nu}\, ,\\
&&Q_s \tau_{\rm grow} \sim 1/\sqrt{\nu}\, .
\EQA
If the waiting 
time $Q_s \tau_{\rm wait}$ plus the growth time $Q_s \tau_{\rm grow}$
is within the validity region (\ref{validity_E}), one may be able to 
conclude the existence of instability. However, this condition reads
\BQ
Q_s(\tau_{\rm wait}+\tau_{\rm grow})\sim 
\left(1 + \frac{1}{\nu}\right)\sqrt{\nu} \ll \sqrt{2\nu}\, ,
\EQ
which is, unfortunately, not satisfied by any value of $\nu>0$. 
Therefore, the approximate equation (\ref{diff_fluc_E2}) allows 
for unstable solution, but it shows the instability only outside 
the region of validity.

Indeed, this means that one cannot ignore the
the $E^2$ term in Eq.~(\ref{diff_fluc_E1}), 
and then one finds that its solution does not show instability.
This is immediately understood if one notices that the second 
term in the curly bracket can be written as $(\nu\pm gE\tau^2/2)^2/\tau^2$
which is always non-negative. 
Thus, the curly bracket in Eq.~(\ref{diff_fluc_E1})
never becomes negative. On the other hand, the instability 
in the approximate equation (\ref{diff_fluc_E2}) occurs 
when the corresponding part becomes negative, which is not 
allowed in the original equation (\ref{diff_fluc_E1}). 
And the upper limit of the time (\ref{validity_E}) corresponds to 
the time when $(\nu- gE\tau^2/2)^2/\tau^2=0$. 

In fact, one can write down the explicit form of the solution 
to the original equation:
\BQ
\tilde b^{(\pm)} \propto \frac{1}{\tau}\left\{ c_+ 
M_{\kappa,\mu}\left(i\frac{gE}{2}\tau^2\right)+
c_- M_{\kappa,-\mu}\left(i\frac{gE}{2}\tau^2\right)\right\}\, ,
\EQ
where $M_{\kappa,\mu}(z)$ is the Whittaker function and 
$\kappa$ and $\mu$ are both pure imaginary numbers
\BQ
\kappa=-\frac{i}{2}\left(\frac{\kk^2}{gE}\pm \nu\right)\, ,\qquad 
\mu = i \frac{\nu}{2}\, .
\EQ
By using the asymptotic form of the Whittaker function, one finds 
that there is no instability at large $\tau$. 

What we have done should be essentially equivalent to 
the Schwinger mechanism. Unlike the creation of massive 
fermion and anti-fermion pairs, there is no mass gap for 
the `charged' gluon matter $\tilde b^{(\pm)}$. Thus any external 
electric field allows pair creations. 
In fact, the `charged' fluctuations are accelerated in the 
electric field, and acquire large longitudinal momentum 
$\nu\pm gE\tau^2/2=\nu \pm g{\mathcal A}_\eta$ \cite{Mottola}. 
Therefore, even if the fluctuation does not grow in time, 
it will be infinitely accelerated towards the ends which are 
moving at the speed of light.
This situation may be also called instability as discussed 
in Ref.~\cite{Chang-Weiss}. 
It should be possible to find more explicit correspondence 
between our case and the Schwinger mechanism, and it will be
interesting to also consider the effect of backreaction.
We leave such topics for future study.

\subsubsection{Time-dependent electric field}

Let us briefly discuss the case with  time-dependent electric field. 
Although we can in principle treat any time-dependent electric 
background, we here discuss only the following simple case where 
we can easily find the solution to the equation:
\BQ
E(\tau)=E_0 \left(\frac{\tau_0}{\tau}\right)^\gamma \, ,\quad 
\gamma\neq 2\, .\label{time_dep_E}
\EQ
To obtain the differential equation for the fluctuations, 
one has to replace $E$ in Eqs.~(\ref{diff_fluc_E1}) and 
(\ref{diff_fluc_E2}) by $2E(\tau)/(2-\gamma)\equiv \beta E(\tau)$ 
(instead of $E(\tau)$). Then, for the approximate equation 
(\ref{diff_fluc_E2}),
\BQ
\left(\del_\tau^2 +\frac{1}{\tau}\del_\tau\right)\tilde b^{(\pm)}
+\left(\kk^2 + 
\frac{\nu^2}{\tau^2} \pm 
\beta \nu gE_0\left(\frac{\tau_0}{\tau}\right)^\gamma 
\right)\tilde b^{(\pm)}=0\, .
\EQ
One can reproduce the time-independent case by setting $\gamma=0$ 
($\alpha=1$).
When $\kk=0$, the solution is again given by the modified Bessel 
function. For example, 
\BQ
\tilde b^{(-)}\propto 
I_{i\beta\nu}
\left(\sqrt{{\beta^3}{\nu g E(\tau)}}\, \tau\right)\, .
\EQ
This solution should be a good approximation for small $\tau$,
but the unstable behavior at large $\tau$ is just an artifact which 
appears when we ignore the $E^2$ term.

\section{Summary and discussion}

We have investigated the dynamics and stability of the glasma 
based on the picture that the glasma is initially made of 
longitudinally extending flux tubes with their transverse size 
being $1/Q_s$. For simple Gaussian flux tubes where the 
longitudinal fields have only one color component, we found 
that the flux tubes expand outwards as time goes, and 
the transverse field strengths are accordingly induced.
This time evolution is quite consistent with the numerical 
result reported in Ref.~\cite{Lappi-Larry}, and thus we 
expect that our simple picture captures the essence of
the actual physics situation. It is very remarkable that the Abelian 
dynamics of the background field in our model reproduces 
the qualitative behavior of the time-dependence of the energy 
density found in Ref.~\cite{Lappi-Larry}. At the same time,
this raises a question what is the role of the non-Abelian nature of QCD 
in the result of Ref.~\cite{Lappi-Larry}. One of the possible 
effects of non-Abelian dynamics will be the interactions 
between flux tubes with different colors. This was ignored 
in our calculation for an isolated flux tube, but we expect 
such effects will become relevant only at later time when 
the overlap between the expanding flux tubes is sizable.

We have studied in detail the response of small rapidity-dependent
fluctuations in the background of boost-invariant glasma. 
Instabilities in the electric or magnetic fields are, if any, 
expected for the `charged' fluctuations since they can couple 
to the background field. Recall that there is no preferred color 
direction in the Yang-Mills theory. 
Thus, if the field is forced to have only a certain color component, 
it is quite natural that fluctuations perpendicular to the 
forced color component (namely, the `charged' fluctuations) 
become large, and can grow exponentially in some case. In fact,
we found that the `charged' fluctuations can grow exponentially 
only in the magnetic background. On the other hand, 
the response of the `charged' fluctuations in the electric 
background is essentially equivalent to the Schwinger mechanism. 
While the `charged' fluctuation is accelerated infinitely 
and the system is unstable in the Schwinger mechanism, 
the fluctuation itself does not grow in time. 

The unstable mode in the magnetic background corresponds 
to the lowest Landau level whose transverse size is $1/Q_s$. 
Thus it can survive in the flux tube which expands in the 
transverse directions with its initial size being $1/Q_s$. 
The time for the instability to become manifest is 
estimated as $Q_s \tau \sim \nu+1$ where $\nu$ is the 
momentum conjugate to the rapidity coordinate. 
Thus, for small $\nu$, instability may occur at very early stage
of the collisions, even before the ordinary plasma 
instability starts to operate. Thus, we hope that the glasma 
instability  play an important role in the early thermalization.

As we have already emphasized, the coupling between the 'charged' 
fluctuation and the magnetic background field is the essential 
dynamics for the existence of instability. Since the 'charged'
fluctuation itself is the gauge field, such coupling is 
possible only in the non-Abelian Yang-Mills theories. Therefore, 
one may say that the glasma instability is a result of 
non-linearity of the Yang-Mills equation. Having said this, 
we recall the similarity of the glasma instability with the 
emergence of chaos in the classical Yang-Mills theory \cite{chaos}. 
It is known that the spatially constant Yang-Mills fields exhibit
chaos due to the non-linear interactions among the fields. 
Notice that if there are unstable modes as in the glasma, 
the Lyapunov exponent for two unstable modes with slightly 
different initial conditions is given by the growth constant of 
the unstable modes. This suggests an interesting viewpoint
of the Yang-Mills chaos to the glasma instability.

The Weibel instability has been widely studied
as the main mechanism
of the plasma instability in the context of heavy ion physics 
\cite{Plasma_instability}.
Although it is not evident at present whether the magnetic instability 
found in the present paper is equivalent to it,
our setup is quite similar to that of the 
plasma instability scenario. For example, the 'hard' particles 
correspond to the boost-invariant background field in our study,
which are supplied by the CGC initial condition, and the 'soft' fields
are the rapidity-dependent fluctuations in our study.
Since the background field is 
independent of rapidity, it yields a very anisotropic distribution
in the momentum space 
($k_\perp\sim Q_s,\ k_z\sim 0$).
Moreover, linearizing the Yang-Mills equations for
the fluctuations $a_\mu$:
$$
K^{\mu\nu}[{\mathcal A}]a_\nu=0,\qquad 
K^{\mu\nu}[{\mathcal A}]=\left. 
\frac{1}{2}\frac{\delta {\mathcal L}_{\rm YM}}{\delta A_\mu\delta A_\nu}
\right\vert_{A_\mu={\mathcal A}_\mu}, 
$$
is nothing but the computation of the inverse propagator
$K^{\mu\nu}[{\mathcal A}]$ of the `soft' fields
on the background field ${\mathcal A}_\mu$.
This may correspond
to the hard-loop calculation in the study of the plasma instability. 
We expect that future investigation will clarify the possible 
relationship between our result and the Weibel instability.

\section*{Acknowledgements}

The authors wish to thank Aiichi Iwazaki for discussions and 
useful correspondences, and Yasuyuki Akiba for urging them to 
clarify the dynamics of expanding flux tubes. They also thank 
Kenji Fukushima, Fran\c{c}ois Gelis, Tuomas Lappi,
Larry McLerran,
and the members of KEK Nuclear-Hadron Theory Group and
Komaba Nuclear Theory Group for their interests in this work.
They are grateful to
the YITP International Symposium ``{\it Fundamental Problems in 
Hot and/or Dense QCD}'' during which the present work was finalized. 
This work is partly supported by Grants-in-Aid
(18740169 (KI) and 19540273 (HF)) of MEXT.


\begin{thebibliography}{99}

\bibitem{CGC}
  For a recent review, see F.~Gelis, T.~Lappi and R.~Venugopalan,
  %``High energy scattering in Quantum Chromodynamics,''
  Int.\ J.\ Mod.\ Phys.\  E {\bf 16} (2007) 2595
  [arXiv:0708.0047 [hep-ph]].

\bibitem{Heinz}
  For example, U.~W.~Heinz and P.~F.~Kolb,
  ``{\it Two RHIC puzzles: Early thermalization and the HBT problem},''
  arXiv:~hep-ph/0204061.

\bibitem{Plasma_instability} 
   For a recent review, see S.~Mrowczynski,
  %``Early stage thermalization via instabilities,''
  PoS C {\bf POD2006} (2006) 042
  [arXiv:hep-ph/0611067].


\bibitem{YKIS}K.~Itakura, Prog. Theor. Phys. Supp. 168 (2007) 295.



\bibitem{Mueller-Son}
  A.~H.~Mueller and D.~T.~Son,
  %``On the equivalence between the Boltzmann equation and classical field
  %theory at large occupation numbers,''
  Phys.\ Lett.\  B {582} (2004), 279, 
  T.~Stockamp,
  %``Classical approximation of the Boltzmann equation in high energy QCD,''
  J.\ Phys.\ G {\bf 32} (2006) 39
  [arXiv:hep-ph/0408206].


\bibitem{Lappi-Larry}T.~Lappi and L.~McLerran, Nucl. Phys. A772 (2006) 200.


\bibitem{Raju}
  P.~Romatschke and R.~Venugopalan,
   Phys.\ Rev.\ Lett.\  {\bf 96} (2006) 062302
  [arXiv:hep-ph/0510121]; 
  Phys.\ Rev.\  D {\bf 74} (2006) 045011
  [arXiv:hep-ph/0605045].


\bibitem{Romatschke-Rebhan} P.~Romatschke and A.~Rebhan,
  %``Plasma instabilities in an anisotropically expanding geometry,''
  Phys.\ Rev.\ Lett.\  {\bf 97} (2006) 252301
  [arXiv:hep-ph/0605064].


\bibitem{Rebhan2}
  A.~Rebhan, M.~Strickland and M.~Attems,
  ``{\it Instabilities of an anisotropically expanding non-Abelian 
   plasma: 1D+3V discretized hard-loop simulations},'' 
  arXiv:0802.1714 [hep-ph].
  %%CITATION = ARXIV:0802.1714;%%





\bibitem{Initial_fluctuation}
  K.~Fukushima, F.~Gelis and L.~McLerran,
  %``Initial singularity of the little bang,''
  Nucl.\ Phys.\  A {\bf 786} (2007) 107
  [arXiv:hep-ph/0610416].


\bibitem{Muller}  S.~G.~Matinyan, B.~Muller and D.~H.~Rischke,
  %``Classical gluon radiation in ultrarelativistic nuclear collisions:
  %Space-time structure, instabilities, and thermalization,''
  Phys.\ Rev.\  C {\bf 57} (1998) 1927
  [arXiv:nucl-th/9708053].


%\cite{Kharzeev:2005iz}
\bibitem{Kharzeev1}
  D.~Kharzeev and K.~Tuchin,
  %``From color glass condensate to quark gluon plasma through the event
  %horizon,''
  Nucl.\ Phys.\  A {\bf 753}, 316 (2005)
  [arXiv:hep-ph/0501234].
  %%CITATION = NUPHA,A753,316;%%


%\cite{Kharzeev:2006zm}
\bibitem{Kharzeev2}
  D.~Kharzeev, E.~Levin and K.~Tuchin,
  %``Multi-particle production and thermalization in high-energy QCD,''
  Phys.\ Rev.\  C {\bf 75}, 044903 (2007)
  [arXiv:hep-ph/0602063].
  %%CITATION = PHRVA,C75,044903;%%


\bibitem{Fukushima}
  K.~Fukushima,
  %``Initial fields and instability in the classical model of the heavy-ion
  %collision,''
  Phys.\ Rev.\  C {\bf 76} (2007) 021902
  [Erratum-ibid.\  C {\bf 77} (2007) 029901]
  [arXiv:0704.3625 [hep-ph]].



\bibitem{Iwazaki1}
  A.~Iwazaki, ``{\it Thermalization of Color Gauge Fields 
in High Energy Heavy Ion Collisions,}'' arXiv:0712.1405 [hep-ph].


\bibitem{Iwazaki2}
  A.~Iwazaki, ``{\it Decay of Color Gauge Fields in 
Heavy Ion Collisions and Nielsen-Olesen Instability}," 
arXiv:0803.0188 [hep-ph].






\bibitem{Glasma_basic}A.~Kovner, L.~McLerran and H.~Weigert, Phys. Rev. 
D {\bf 52} (1995) 6231, {\it ibid.}  Phys. Rev. 
D {\bf 52} (1995) 3809. 



\bibitem{color_neutral}E.~Iancu, K.~Itakura and L.~McLerran,
  %``A Gaussian effective theory for gluon saturation,''
  Nucl.\ Phys.\  A {\bf 724} (2003) 181
  [arXiv:hep-ph/0212123].


\bibitem{Savvidy}I.~A.~Batalin, S.~G.~Matinyan and G.~K.~Savvidy,
  %``Vacuum Polarization By A Source-Free Gauge Field,''
  Sov.\ J.\ Nucl.\ Phys.\  {\bf 26} (1977) 214
  [Yad.\ Fiz.\  {\bf 26} (1977) 407],
M.~Gyulassy and A.~Iwazaki,
  %``Quark And Gluon Pair Production In SU(N) Covariant Constant Fields,''
  Phys.\ Lett.\  {\bf 165B} (1985) 157.






\bibitem{tau-exp}R.~J.~Fries, J.~I.~Kapusta, and Y.~Li, "{\it Near-Fields and 
Initial Energy Density in High Energy Nuclear Collisions}", nucl-th/0604054.


\bibitem{Gyulassy-Matsui}
  M.~Gyulassy and T.~Matsui,
  %``Quark Gluon Plasma Evolution In Scaling Hydrodynamics,''
  Phys.\ Rev.\  D {\bf 29} (1984) 419.
  %%CITATION = PHRVA,D29,419;%%




\bibitem{Chang-Weiss}S.~J.~Chang and N.~Weiss, Phys. Rev. D20 (1979) 869.  

\bibitem{Nielsen-Olesen}
  N.~K.~Nielsen and P.~Olesen,
  %``An Unstable Yang-Mills Field Mode,''
  Nucl.\ Phys.\  B {\bf 144} (1978) 376.

\bibitem{Mottola}F.~Cooper, J.~M.~Eisenberg, 
Y.~Kluger, E.~Mottola and B.~Svetitsky,
  %``Particle production in the central rapidity region,''
  Phys.\ Rev.\  D {\bf 48} (1993) 190
  [arXiv:hep-ph/9212206].

\bibitem{chaos}
  T.~S.~Biro, S.~G.~Matinyan and B.~Muller,
  ``{\it Chaos and gauge field theory},'' 
  World Sci.\ Lect.\ Notes Phys.\  {\bf 56} (1994) 1.




\end{thebibliography}
\end{document}